\documentclass[showpacs,aps,prb,twocolumn,superscriptaddress]{revtex4}
\usepackage{amsmath}
\usepackage{graphicx} 
\usepackage{dcolumn}
\usepackage{bm}
\usepackage{amssymb,amsmath}

\usepackage{multirow}
\usepackage{natbib}

\begin{document}

\title{Above-Bandgap Magneto-optical Kerr Effect in Ferromagnetic GaMnAs}

\author{C. Sun}
\author{J. Kono}
\email[]{kono@rice.edu}
\thanks{corresponding author.}
\affiliation{Department of Electrical and Computer Engineering, Rice University, Houston, Texas 77005, USA}
\affiliation{Department of Physics and Astronomy, Rice University, Houston, Texas 77005, USA}
\affiliation{The Richard E. Smalley Institute for Nanoscale Science and Technology, Rice University, Houston, Texas 77005, USA}

\author{Y.-H. Cho}
\author{A. K. W\'{o}jcik}
\author{A. Belyanin}
\affiliation{Department of Physics, Texas A\&M University, College Station, Texas 77843, USA}

\author{H. Munekata}
\affiliation{Imaging Science and Engineering Laboratory, Tokyo Institute of Technology, Yokohama, Kanagawa
226-8503, Japan}

\date{\today}

\begin{abstract}
We have performed a systematic magneto-optical Kerr spectroscopy
study of GaMnAs with varying Mn densities as a function of
temperature, magnetic field, and photon energy. Unlike previous
studies, the magnetization easy axis was perpendicular to the sample
surface, allowing us to take remanent polar Kerr spectra in the
absence of an external magnetic field.  The remanent Kerr angle
strongly depended on the photon energy, exhibiting a large positive
peak at $\sim1.7$~eV. This peak increased in intensity and
blue-shifted with Mn doping and further blue-shifted with annealing.
Using a 30-band ${\bf k\cdot p}$ model with
antiferromagnetic $s,p$-$d$ exchange interaction, we calculated the dielectric tensor of GaMnAs in the interband transition region, assuming that our samples are in the metallic regime  and the impurity band has merged with the valence band. We successfully
reproduced the observed spectra without \emph{ad hoc}
introduction of the optical transitions originated from impurity states in the band gap.  These results lead us to
conclude that above-bandgap magneto-optical Kerr rotation in
ferromagnetic GaMnAs is predominantly determined by interband transitions between the conduction and valence bands.
\end{abstract}


\maketitle


\section{Introduction}

Dilute magnetic semiconductors,\cite{SCSM88} i.e., II-VI or
III-V semiconductors doped with transition metal ions, show a
variety of magnetic phenomena arising from the interplay between
localized and delocalized carriers.  The strong $s,p$-$d$ exchange
interaction between band carriers ($s$,$p$) and Mn local moments
($d$) results in enormous enhancement of carrier $g$-factors as well
as the formation of magnetic polarons.  The discovery of
carrier-induced ferromagnetism in (III,Mn)V
semiconductors\cite{Munekata89.1849, Ohno92.2664, Munekata93.2929,
Ohno96.363} with relatively high Curie temperatures ($T_{\rm c} <$ 190~K)\cite{NovaketAl08PRL}
has stimulated further interest in these systems both from applied
and fundamental viewpoints.\cite{MacDonaldetAl05NatMat} Although it
has been experimentally established that the ferromagnetic
interaction between Mn moments is mediated by free
holes,\cite{Koshihara97.4617, Ohno00.944} there remain some basic
questions as to the nature of the free holes ($p$-like or $d$-like)
and the role of Mn impurity bands in transport and optical
processes.

Magneto-optical (MO) spectroscopy is useful for studying
spin-dependent electronic states in magnetic systems via
polarization-dependent reflection and absorption, i.e., the
MO Kerr effect (MOKE) and magnetic circular dichroism (MCD).
Previous MO studies of GaMnAs,\cite{AndoetAl1998JAP,
SzczytkoetAl1999PRB, BeschotenetAl1999PRL, OkabayashietAl2001PRB, KomorietAl2003PRB, SingleyetAl2003PRB, BurchetAl2004PRB, BurchetAl2006PRL,
LangetAl2005PRB, ChakarvortyetAl2007APL, AndoetAl2008PRL,
BerciuetAl2009PRL, AcbasetAl2009PRL} however, have produced much
controversy regarding
 the value and sign of
the $p$-$d$ exchange coupling constant, $J_{pd}$, the position of the Fermi level, the insulating vs. metallic nature of the samples, and the contribution of the optical transitions originated from Mn impurity states.

One of the commonly observed characteristics in MO spectra for ferromagnetic GaMnAs is that the sign of MO signal above the band gap ($\sim1.4$ to $\sim2$~eV) corresponds to greater absorption for $\sigma^-$ polarization of light than for $\sigma^+$ polarization, which is opposite to (II,Mn)VI systems\cite{SCSM88} as well as paramagnetic GaMnAs with low Mn doping. An opposite sign of MO signal
could indicate an opposite sign of exchange coupling.  Szczytko {\it
et al}.\cite{SzczytkoetAl1999PRB} proposed that the sign difference
is a result of $E_{\rm F}$ being inside the valence band in ferromagnetic GaMnAs
while $p$-$d$ exchange coupling is still antiferromagnetic ($J_{pd}>
0$) as in (II,Mn)VI; due to exchange splitting, the Moss-Burstein
shift~\cite{MossetAl1954PPS} is expected to become
spin-dependent, and thus, the lowest-energy $\sigma^+$ transition
should occur at a large momentum, while the lowest-energy $\sigma^-$
transition should still occur near the zone center. Komori {\it et
al}.~also explained their MO data by adopting this
model.\cite{KomorietAl2003PRB} However, Lang {\it et
al}.,\cite{LangetAl2005PRB}~in analyzing their MO data within a
parabolic band model assuming that $E_{\rm F}$ resides inside the valence band, had
to conclude that $p$-$d$ exchange coupling is ferromagnetic ($J_{pd}
< 0$); in order to explain their data quantitatively, they were also
forced to introduce a dispersionless level of unclear origin inside
the conduction band that would need to have a very large oscillator strength
(three times larger than the usual valence-band-to-conduction-band transitions). In order
to identify the optical contribution from the impurity band, infrared
spectroscopy studies have been carried out as it probes electronic
states near the Fermi surface. Burch {\it et
al}.\cite{BurchetAl2004PRB, BurchetAl2006PRL}~interpreted the
observed peak around 0.2~eV as the valence-band-to-impurity-band transition.  On the
contrary, Acbas {\it et al}.\cite{AcbasetAl2009PRL}~measured
infrared Kerr and Faraday effects in GaMnAs and explained the
spectra with a valence-band-hole theory.  In addition, Ando {\it et
al}.\cite{AndoetAl2008PRL}~and Berciu {\it et
al}.\cite{BerciuetAl2009PRL}~both performed a MCD study on
paramagnetic and low-$T_c$ ferromagnetic GaMnAs samples. Ando {\it
et al}.~claimed that the MO features below 1.4~eV are likely to
be related to impurity states.  They interpreted their
above-band-gap spectra in ferromagnetic samples as being composed
of a broad positive impurity-related background and negative and
positive peaks associated with band-edge singularities ($E_0$ and
$E_0 + \Delta_0$ critical points, respectively), concluding that
$E_{\rm F}$ is in the Mn impurity band and there is no Moss-Burstein shift.  Berciu {\it et
al}., based on their sample-dependent analysis and simplified one-dimensional model, proposed a unified
interpretation of MCD spectra taking into account both impurity band and valence band
contributions. Different from Ando {\it et al}., they concluded that
the broad positive feature was due to valence-band-to-conduction-band transitions while the
negative MCD near 1.4~eV, which occurred only when the sample was
fully magnetized, was due to impurity-band-related transitions.

\begin{figure}
\includegraphics [scale=0.55] {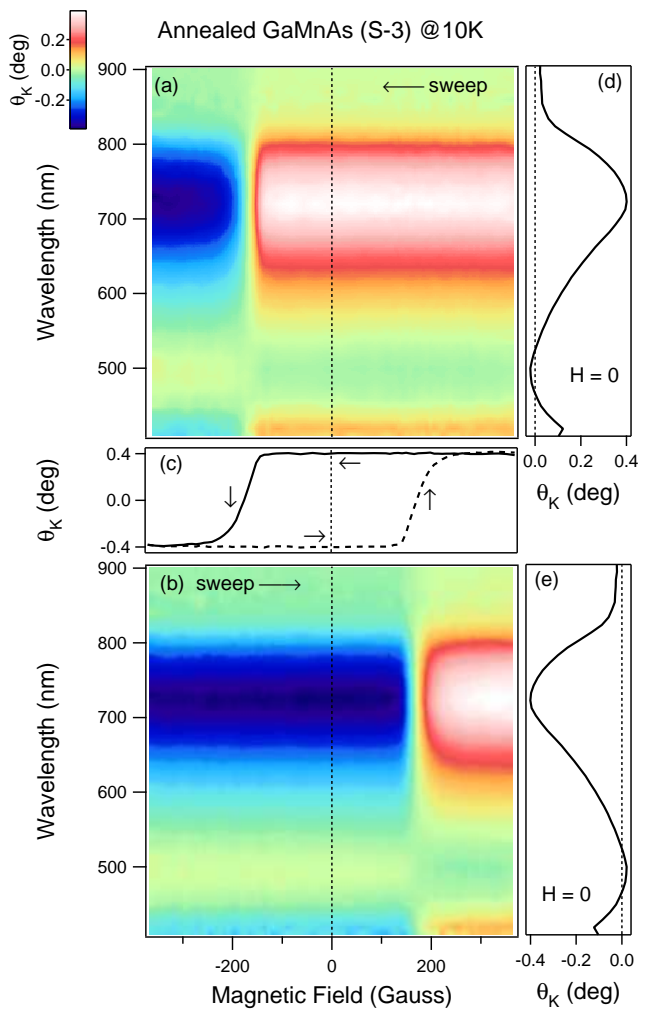}
\caption{Contour plot of the measured Kerr angle as a function of
magnetic field and wavelength for an annealed
Ga$_{0.976}$Mn$_{0.024}$As (S-3) at 10~K. Magnetic field was swept
from $+400$~Gauss to $-400$~Gauss in (a) and reversed in (b). (c) A
hysteresis loop with magnetic field swept from $+400$ to
$-400$~Gauss (solid line) and from $-400$ to $+400$~Gauss (dashed
line). (d) and (e) show wavelength dependence of remanent Kerr angle
(i.e., no external magnetic field) for down-sweep and
up-sweep, respectively.} \label{3D10K}
\end{figure}
%
Here, we report results of a systematic MOKE spectroscopy study of
GaMnAs samples with different Mn densities with three continuously
varying experimental parameters: temperature (10~K - 75~K), magnetic
field ($-400$ - $+400$~Gauss), and photon energy (1.4~eV - 3.1~eV).
Figure~\ref{3D10K} displays representative Kerr rotation data
measured for a Ga$_{0.976}$Mn$_{0.024}$As sample (S-3) at 10~K when
the magnetic field was swept between positive 400~Gauss and negative
400~Gauss. Strong variations of the Kerr angle are seen as a
function of wavelength and magnetic field.  Magnetic hysteresis is
also evident by comparing the down-sweep [in (a)] and up-sweep
[in (b)] data.

We focus on the \textit{remanent} Kerr angle, i.e., the Kerr angle at zero external magnetic field, which was found to strongly depend on the photon energy, exhibiting
positive peaks near 1.7~eV and 3~eV and a negative peak near 2.5~eV.
Without an external magnetic field, the MOKE spectra were caused by
the interband transitions between the spontaneously spin-split bands. The
$\sim1.7$~eV peak increased in intensity and blue-shifted with the
Mn density and further blue-shifted with annealing. Using a 30-band ${\bf k\cdot p}$ model with
antiferromagnetic sign of p-d exchange interaction, disorder broadening, and
density dependent band gap, we calculated the dielectric tensor of GaMnAs in the interband transition region and successfully reproduced the observed
spectra assuming that our samples are in the metallic regime  and the Mn impurity band has merged with the valence band. There was no need to add any background dielectric constant or the optical transitions
involving impurity states located in the band gap in order to explain our
spectra.

\section{Samples and Experimental Methods}\label{sec:Exp}
%
\begin{table}
\begin{center}
\begin{tabular}{c c c c}
\hline\hline
sample&\hspace{0.9cm}S-1&\hspace{0.9cm}S-2&\hspace{0.9cm}S-3\\$x$ &
\hspace{0.9cm}0.01 & \hspace{0.9cm}0.024 & \hspace{0.9cm}0.024
\\$T_{\rm c}$(K) &
\hspace{0.9cm}30 & \hspace{0.9cm}45 & \hspace{0.9cm}70\\
&\hspace{0.9cm}unannealed&\hspace{0.9cm}unannealed&\hspace{0.9cm}annealed\\
\hline
\end{tabular}
\end{center}
\caption{Ga$_{1-x}$Mn$_{x}$As samples studied in this paper. S-3 is
annealed and the other two are not.} \label{samples}
\end{table}
\begin{figure}
\includegraphics [scale=0.55] {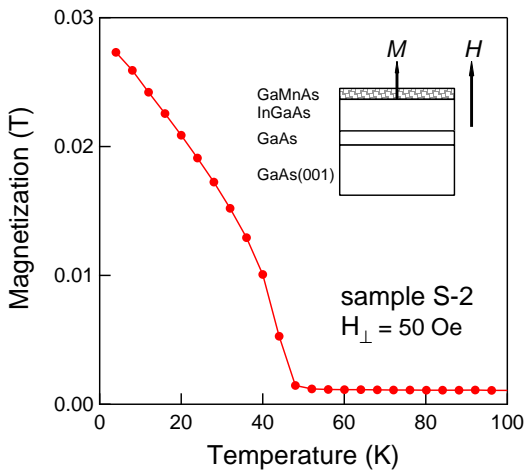}
\caption{Temperature dependence of magnetization for sample S-2.  A
small magnetic field of 50~Oe was applied perpendicular to the
sample surface.  $T_{\rm c} = 45$~K is revealed. Inset: sample structure.}
\label{SQUID}
\end{figure}
\begin{figure}
\includegraphics [scale=0.58] {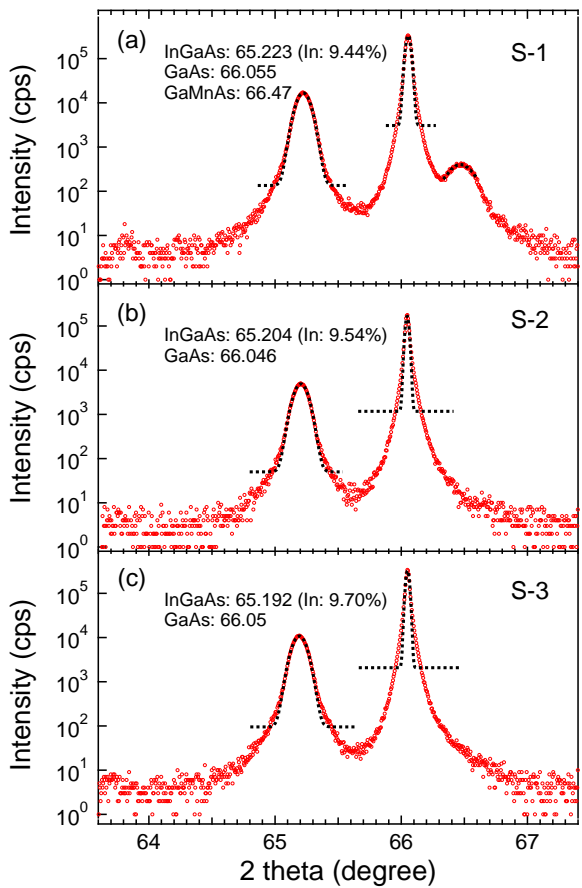}
\caption{X-ray diffraction patterns for three GaMnAs samples: (a)
S-1, (b) S-2, and (c) S-3. Peaks due to the diffraction from the
(004) planes are measured. InGaAs and GaAs peaks are observed around
65.2 and 66 degrees for all three samples.  However, the GaMnAs peak is
only observed in S-1 around 66.5 degrees.} \label{XRD}
\end{figure}
The three GaMnAs samples studied were grown by low-temperature
molecular beam epitaxy (LT-MBE). They have a similar structure
consisting of a 50-nm epilayer of Ga$_{1-x}$Mn$_{x}$As and a buffer
layer of 1000-nm In$_{0.14}$Ga$_{0.86}$As on a GaAs (001) substrate.
The sample structure is depicted in the inset of Fig.~\ref{SQUID}.
Sample S-1 had a nominal Mn content ($x$) of 0.01 and $T_{\rm c}$ = 30~K.
Sample S-2 had a nominal Mn content of 0.024 and $T_{\rm c}$ = 45~K. The
third sample, S-3, was from the same wafer as S-2, but was further
annealed in air at 190$^{\circ}\mathrm{C}$ for four hours. Its $T_{\rm c}$
was increased to 70~K.  Annealing is known to remove the
interstitial Mn defects (Mn$_{\textrm{i}}$) in LT-MBE GaMnAs and
increase the effective Mn content $x$
and the hole density $p$.\cite{BlinowskietAl2003PRB, EdmondsPRL2004, Zhao} As a donor,
Mn$_{\textrm{i}}$ compensates acceptors and decreases the hole
carrier concentration in the sample. Mn$_{\textrm{i}}$ also prefers
to be antiferromagnetically coupled to substitutional
Mn$_{\textrm{sub}}$ in such a way that it reduces the local spins in
the lattice. Thus, we expect that sample S-3 has a higher
concentration of holes as well as effective Mn$_{\textrm{eff}}$ than
S-2.  The sample parameters are summarized in
Table~\ref{samples}. All the samples exhibited an ``out-of-plane''
easy axis, i.e., the magnetization direction was
perpendicular to the sample surface. Temperature dependent
magnetization of sample S-2 measured with a small perpendicular
field of 50~Oe is shown in Fig.~\ref{SQUID}. A $T_{\rm c}$ of 45~K is
clearly observed.

The ``out-of-plane'' anisotropy indicates that all three samples are
strained. The x-ray diffraction (XRD) patterns of the three samples
are shown in Fig.~\ref{XRD}. The spacings of the (004) planes of the
InGaAs, GaAs, and GaMnAs layers are measured. In the case of sample
S-1, the peaks of the three layers are observed at 65.2, 66, and
66.5 degrees, respectively, indicating that the lattice parameter of
InGaAs is larger than that of the GaAs, whereas the lattice
parameter of the GaMnAs is smaller than that of the GaAs. Therefore,
the epilayer of sample S-1 is fully strained. The indium composition
inferred from the lattice parameter is about 9.44\% in the buffer
layer. In sample S-2 and S-3, peaks for InGaAs and GaAs remain
around 65.2 and 66 degrees, respectively. However, the peak for
GaMnAs is missing. It indicates that there is no well-defined
lattice parameters for GaMnAs in these two samples. This happens
when the epilayer is neither fully strained nor fully relaxed.
Instead, the lattice parameter changes continuously across the
growth direction. Thus, diffracted x-ray waves do not result in the
well-defined Laue function with a sharp feature.

\begin{figure}
\includegraphics [scale=0.48] {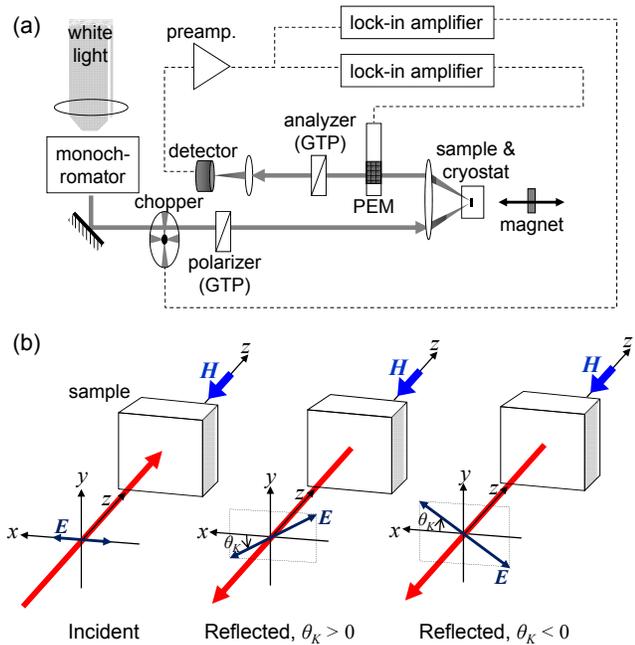}
\caption{(Color online) (a) Schematic diagram of the experimental setup. Light with a selected wavelength is linearly polarized, and quasi-normal incident onto the sample.  The reflected beam from the sample passes through a photoelastic modulator (PEM) and an analyzer, and it is detected by a photodiode.  The signal is amplified and demodulated by two lock-in amplifiers.  The sample is mounted inside a cryostat.  Magnetic field is swept within $\pm$400~Gauss.  (b) Schematic diagrams showing the directions of light propagation, magnetic field, and Kerr rotation.  The incident beam propagates in the positive $z$-direction and is linearly polarized in the $x$-direction.  The reflected beam travels in the negative $z$-direction.  It becomes elliptically polarized and acquires a Kerr rotation $\theta_{\rm K}$ away from the $x$-direction as shown in the right two panels in (b).  $\theta_{\rm K}$ is defined to be positive (negative) if the rotation is clockwise (counterclockwise) when looking in the positive z-direction.
} \label{setup}
\end{figure}
%
MOKE measurements were performed in the polar geometry in which the light beam was quasi-normally incident on the sample surface.  Figure~\ref{setup}(a) is a schematic diagram of the experimental setup.  White light from a 100~W Xe lamp is first focused into a monochromator.  Light with selected wavelength is then polarizaed with a Glan-Thompson polarizer and modulated with a mechanical chopper.  The linearly polarized light impinges on the magnetic sample at a nearly normal incident angle.  The reflected beam passes through a photoelastic modulator (PEM), and an analyzer, and then its intensity is detected with a Si photodiode.  The current signal is amplified and converted into a voltage and fed into lock-in amplifiers.  The Kerr angle caused by the different reflection of $\sigma^\pm$ is proportional to the signal at the second harmonic of the PEM modulation frequency.  The two lock-in amplifiers are used to demodulate the signal.  The first lock-in amplifier is referenced to the chopper frequency to provide a measurement of the average light intensity at each wavelength.  The second lock-in amplifier is referenced to the second harmonic of the PEM frequency to record the fast oscillating signal at 100~kHz.  The Kerr rotation angle ($\theta_{\rm K}$) is derived from the ratio of these two.

Figure~\ref{setup}(b) schematically shows our configuration in more detail.  Note that the magnetic field vector is opposite to the wave vector of the incident light.  This is important in determining how the refractive indices of circularly polarized normal waves in the material are related to the components of the dielectric tensor. The Kerr angle is defined as an arctangent of the negative ratio of the $y$-component to the $x$-component of the reflected electric field, assuming that the incident wave propagating in the positive $z$-direction is linearly polarized in the $x$-direction.  To determine the sign of rotation, we first replace the sample with a silver mirror and induce a rotation of polarization by rotating the polarizer in front of the cryostat.  Depending on the direction of rotation (clockwise or counterclockwise), there is a $\pi$-phase shift in the lock-in reading.  We define it to be positive if the rotation is clockwise.  When the sample is mounted and a magnetic field is applied in the negative $z$-direction, the polarization rotation, i.e., the Kerr angle, is then measured and recorded.

Samples are kept in a helium-flow cryostat, allowing us to vary
the temperature ($T$) from 10~K to above $T_{\rm c}$. An external magnetic
field ($H$) is applied perpendicular to the sample surface. The coercivity
and saturation field are known to be small for all three samples
from SQUID measurements. Thus, the magnetic field is swept within
the range between $\pm$400~Gauss. The Faraday rotation induced by
the cryostat window was subtracted. Any polarization anisotropy
caused by components in the setup was carefully calibrated and
subtracted to get the accurate Kerr rotation.

\section{Experimental Results}
\begin{figure}
\includegraphics [scale=0.55] {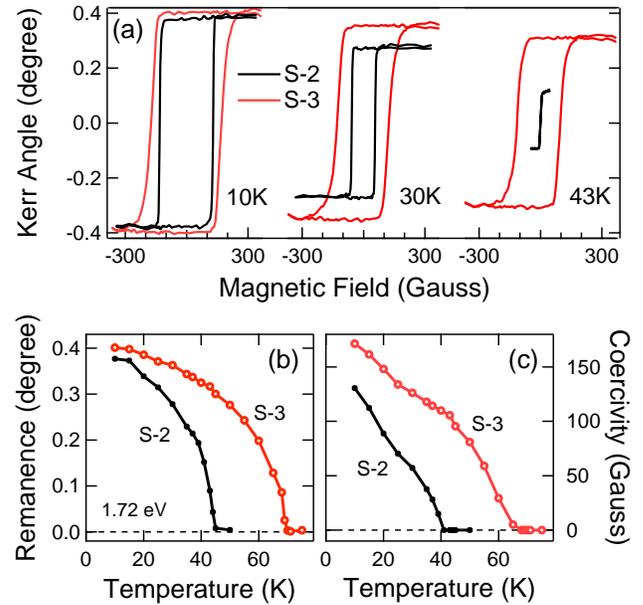}
\caption{(color online). (a)Hysteresis loops (720~nm) detected for
sample S-2 and S-3 at 10~K, 30~K and 43~K. Annealed sample S-3
exhibits larger saturation Kerr angle and coercive field at all
temperatures. Temperature dependence of the (b) remanent Kerr angle
and (c) coercivity for S-2 and S-3.} \label{Anneal}
\end{figure}
Hysteresis loops measured with 720~nm (1.72~eV) for sample S-2 and
S-3 are compared in Fig.~\ref{Anneal}(a) at three temperatures,
10~K, 30~K, and 43~K.  Generally, the annealed sample S-3 has a
larger hysteresis loop than S-2 with both larger saturation Kerr
angle and increased coercive field $H_{\rm c}$.  At low temperature
(10~K), the difference between the two is relatively small; the
remanent Kerr angles are 0.4$^\circ$ and 0.38$^\circ$, respectively, and
the coercive fields are $\sim170$~Gauss and $\sim130$~Gauss,
respectively. With increasing temperature, the difference becomes
more and more pronounced. The sharp vertical switching of Kerr angle
near the coercive field indicates the high quality of the sample
with strong perpendicular magnetization for S-2. Annealed sample S-3
also presents strong anisotropy out of plane although the change
over $H_{\rm c}$ occurs at a slightly slower rate when compared to S-2.
Figures \ref{Anneal}(b) and \ref{Anneal}(c) display the temperature
dependence of remanent $\theta_{\rm K}$ (at 1.72~eV) and $H_{\rm c}$,
respectively, for S-2 and S-3.  It is clearly seen that $T_{\rm c}$
increases as a result of annealing.

\begin{figure}
\includegraphics [scale=0.55] {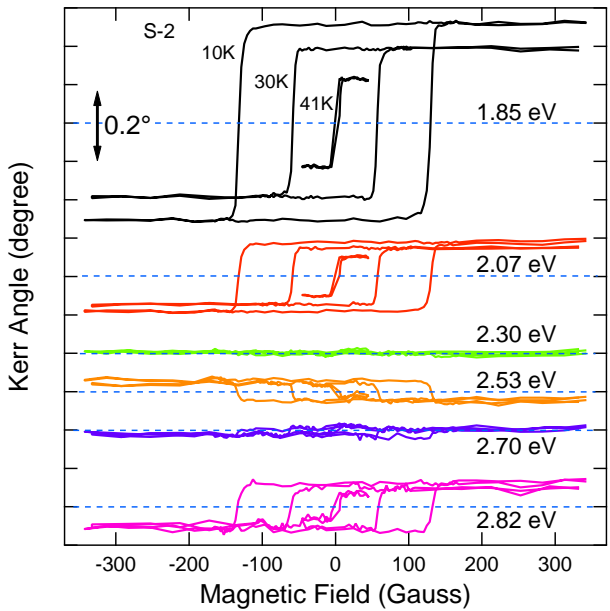}
\caption{(color online).  Temperature dependent hysteresis loops
detected with different photon energies for sample S-2.  Both the
magnitude and sign of the remanent Kerr angle change with photon
energy.  For clarity, the hysteresis loops are vertically offset.}
\label{temperature}
\end{figure}
The photon energy dependent hysteresis loops, $\theta_{\rm K}$ versus $H$,
for sample S-2 are presented in Fig.~\ref{temperature}. For each
photon energy, three hysteresis loops measured below $T_{\rm c}$ ($T$ =
10~K, 30~K, and 41~K) are displayed, and the size of the loop is
seen to shrink with increasing temperature. At each temperature, the
vertical size of the hysteresis loop (i.e., the remanent
$\theta_{\rm K}$) shows considerable variation with photon energy, while
its horizontal width (i.e., the coercivity $H_{\rm c}$) remains
constant. As we can see, the hysteresis loops are sharp and
well-defined at most photon energies but disappear at 2.30~eV and
2.70~eV, although the temperature is below $T_{\rm c}$. In addition, the
sign of remanent $\theta_{\rm K}$ depends on the photon energy.  With the
same sweeping direction, the hysteresis loop at 2.53~eV varies in
the opposite way to that at 2.07~eV and 2.82~eV.

\begin{figure}
\includegraphics [scale=0.58] {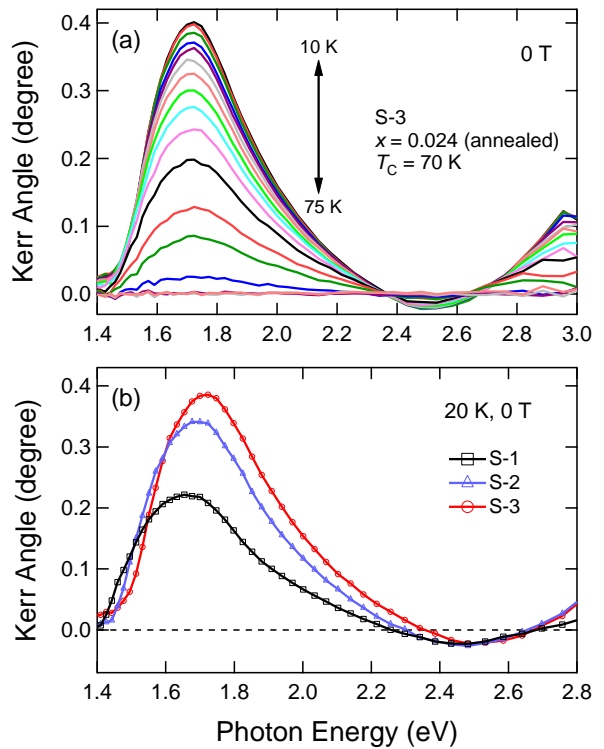}
\caption{(color online).  (a) Remanent ($H$ = 0) Kerr spectra at
different temperatures for S-3.  (b) Remanent Kerr spectra at 20~K
for S-1, S-2, and S-3.  The main peak at 1.7~eV shifts to higher
energy with increasing Mn content and annealing.} \label{KerrSpec}
\end{figure}
Remanent Kerr spectra ($\theta_{\rm K}$ at $H$ = 0 versus photon energy)
for sample S-3 at various $T$ (from 10~K to 75~K) are shown in
Fig.~\ref{KerrSpec}(a).  Above $T_{\rm c}$, there is no Kerr signal left.
Below $T_{\rm c}$, each spectrum contains a pronounced positive peak at
$\sim1.72$~eV and a weaker positive peak at $\sim3$~eV. The spectrum
flips signs in the range of 2.4 - 2.7~eV. With increasing
temperature, the main 1.72~eV peak remains in the same spectral
position, while the magnitude decreases and finally disappears above
70~K. Spectra for S-1 and S-2 were essentially the same as S-3,
except that the main peaks occurred at slightly lower energies.
Spectra for all three samples at 20~K are plotted together in
Fig.~\ref{KerrSpec}(b).  We clearly see that the main peak shifts
from 1.65~eV in S-1 to 1.7~eV in S-2 and to 1.72~eV in S-3, due to
increasing Mn content (S-1 to S-2 and S-2 to S-3) and annealing (S-2
to S-3).

\section{Theoretical modeling}\label{theory}

\subsection{Band Structure Calculation}\label{band-structure}

Our theoretical model is based on the \mbox{$\bf k\cdot p$} method\cite{Luttinger1955,Bir,RichardetAl2004PRB,Radhia} and treats the antiferromagnetic exchange coupling between the hole spins in the host semiconductor and the substitutional Mn$^{2+}$ spins within the mean-field approximation.\cite{Dietl1997,JungwirthetAl1999PRB,Dietletal00Science, Dietl2001, Konig} Recently, this approach was used to explain observed MO effects in GaMnAs in the metallic regime, especially in the infrared spectral region near the Fermi level located inside the valence band.\cite{AcbasetAl2009PRL}  Here we model our experimental magneto-optical Kerr spectra in the interband region with a similar approach, but using a full band structure obtained with a 30-band $\bf k\cdot p$ method\cite{RichardetAl2004PRB} instead of the effective band structure calculation based on the 2-level model, which
includes only $\Gamma_{\rm 6C}$ and $\Gamma_{\rm 8V}$/$\Gamma_{\rm 7V}$ (here we follow the $T_d$ double group notation throughout the paper).  We show that this approach provides quantitatively and qualitatively reasonable results in reproducing the measured Kerr spectra and magnetization.

To model MOKE spectra up to $\sim3$~eV, the extension to the full band structure is critical, since the 8-band $\bf k\cdot p$ method based on the 2-level model, which is often employed to explain the electronic band structure of metallic GaMnAs, is valid only up to $k\sim0.1$~\AA$^{-1}$ measured from the $\Gamma$ edge,  in both the conduction band and the valence band.  Optical transitions with energy $\sim3$~eV involve states with $k$-vectors higher than the above value, where the 2-level model breaks down.  Furthermore, the dielectric tensor in the spectral range of interest obtains significant contributions from transitions with energies higher than 3~eV.  Simulations based on the 8-band $\bf k\cdot p$ model have to include an adjustable static dielectric constant added by hand.

Furthermore, within the 2-level $\bf k\cdot p$ model, one needs to specify the Luttinger parameters in order to take into account the interaction of the valence band with remote bands.  The 30-band Hamiltonian does not contain the Luttinger parameters; instead, the coupling between bands is described in terms of the matrix elements and band edges that are retrieved from the experiments. Once the band structure is determined with the 30-band method, one can calculate the Luttinger parameters\cite{RichardetAl2004PRB, Radhia}  for the subsequent use in simplified 8-band calculations. This is an important advantage of the 30-band $\bf k\cdot p$ method as compared to the 8-band $\bf k\cdot p$ method. Therefore, the modification of the valence band interaction with remote bands due to antiferromagnetic Mn-hole spin exchange coupling, strain, and Coulomb interaction effects in diluted magnetic semiconductors can be simultaneously taken into account within the 30-band $\bf k\cdot p$ model.  Also, the lack of inversion symmetry is taken into account through the matrix element between $\Gamma_{\rm 6C}$ and $\Gamma_{\rm 8V}$/$\Gamma_{\rm 7V}$.

The extended $\bf k\cdot p$ method is known to be valid up to 5~eV above and 6~eV below the top of the valence band, covering interband transitions of energies up to 11~eV.  It has been successfully applied to calculating effective masses and $g$ factors for the group IV and III-V semiconductors.\cite{RichardetAl2004PRB, Fraj}   The general shortcomings of
the method are also known; e.g., limited experimental data for higher conduction bands, the numerical difficulty in ensuring the continuity between U and K symmetry points. The accuracy of the GaAs band structure calculations with the 30-band method is discussed, e.g., in Ref.~\cite{RichardetAl2004PRB}. Here, we do not reproduce all the
complicated basis orbitals, energy level structures, band edge energies, and matrix elements for pure GaAs in a 30-band model, instead referring the reader to Refs.~\onlinecite{RichardetAl2004PRB} and \onlinecite{Radhia}.

Our model Hamiltonian matrix is expressed as
\begin{eqnarray}\label{eq1}
H^{ij}_T& = & H^{ij}_0 + H^{ij}_{\rm strain} + H^{ij}_{\rm spin-ex} \nonumber\\
H^{ij}_{\rm strain}& = & \sum_{\alpha, \beta}\langle
i|D^{\alpha\beta}|j\rangle\varepsilon_{\alpha\beta}^{\rm strain} \\
H^{ij}_{\rm spin-ex}& = & \sum^{N_{Mn}}_{I}\langle
i|J(\mathbf{r}-\mathbf{R}_{I})\mathbf{S}_{I}\cdot\mathbf{s}_{sc}|j\rangle \nonumber\\
               &\approx& J_{ij}N_{Mn}S_{Mn}\langle i|\mathbf{\hat{s}}_{sc}|j\rangle
\nonumber
\end{eqnarray}
where
\begin{equation}
\langle i|D^{\alpha\beta}|j\rangle=-\frac{(p_{\alpha}p_{\beta})_{ij}}{m}+V^{\alpha\beta}_{ij}.
\end{equation}
$H^{ij}_{0}$ is a 30-band $\bf k\cdot p$ matrix, $H^{ij}_{\rm strain}$ is the strain Hamiltonian matrix, $\varepsilon_{\alpha\beta}^{\rm strain}$ is the strain tensor, in which all off-diagonal elements are zero; $\alpha$ and $\beta$ are coordinates, $\langle i|D^{\alpha\beta}|j\rangle$ consists of the linear combination of the deformation
potentials,\cite{Bir} $m$ is the bare electron mass, and $H^{ij}_{\rm spin-ex}$ describes the spin exchange interaction between the substitutional Mn magnetic impurity spins and the itinerant charge carrier spins in the host semiconductor within the mean-field approximation.  The indices $i$ and $j$ run over all 30-band basis orbitals.  The antiferromagnetic spin exchange constant $J_{ij}$ is not zero only when the bases correspond to $\Gamma_{\rm 6C}$, $\Gamma_{\rm 8V}$, and $\Gamma_{\rm 7V}$; $J_{ij} = 54$~meV$\cdot$nm$^3$ for $\Gamma_{\rm 8V}/\Gamma_{\rm 7V}$, $J_{ij} = -9$~meV$\cdot$nm$^3$ for $\Gamma_{\rm 6C}$,\cite{Okabayashi, Szczytko} and $N_{\rm Mn} = 4/a^{3}_{\rm GaMnAs}$, where $a_{\rm GaMnAs}$ is the lattice constant of GaMnAs.  The strain effects are included  only for the bases $\Gamma_{\rm 6C}$, $\Gamma_{\rm 8V}$, and $\Gamma_{\rm 7V}$.  In this approximation, the $\Gamma$ edges of $\Gamma_{\rm 6C}$, $\Gamma_{\rm 8V}$, and $\Gamma_{\rm 7V}$ states are shifted by strain in the same way as those in the 8-band model.

We qualitatively took into account many-body Coulomb interactions through the phenomenological band gap narrowing (BGN).  The hole-occupied exchange spin-split $\Gamma_{\rm 8V}$ bands were assumed to be rigidly shifted by an amount $ap^{1/3}$, where $a = 2.6 \times 10^{-8}$~cm$\cdot$eV and $p$ is in cm$^{-3}$.\cite{JainetAL1990JAP} The proportionality constant is compatible with that used in Ref. \onlinecite{AcbasetAl2009PRL}. Several iterations were necessary to obtain self-consistent positions of the $\Gamma_{\rm 8V}$ edges and the Fermi level.  Still, the resulting BGN shifts are only qualitative; rigorous calculations will be attempted elsewhere.  The disorder effect was also phenomenologically described as broadening of interband optical transitions ($\sim100$~meV half-width at half-maximum) in the linear dielectric response function.

The thermal fluctuations of Mn spin ordering at temperature $T$ = 20~K, at which the experimental data in Fig.~\ref{KerrSpec}(b) was obtained, can be roughly estimated by comparing the measured amplitudes of the remanent Kerr angle as a function of an external magnetic field at 10~K and 20~K.  The effect of thermal fluctuations was included in the calculation of temperature-dependent dielectric tensors and Kerr angle spectra.  The temperature dependence of other physical parameters, for example, the band gap, the lattice constant, the strain tensor, and the hole density were assumed to be negligible in this narrow temperature range.

\begin{figure}
\includegraphics [scale=0.45] {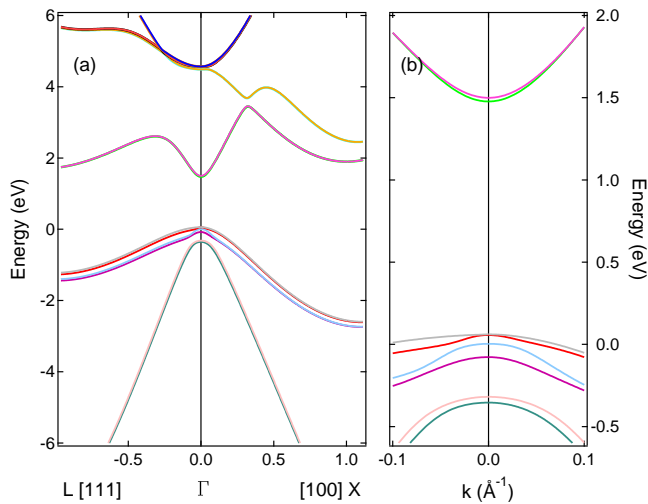}
\caption{(color online).  Band structure of Ga$_{0.95}$Mn$_{0.05}$As, calculated through the 30-band $\bf k\cdot p$ method described in the text, taking into account the Mn-hole spin exchange interaction as well as a biaxial tensile strain of $\varepsilon_{xx}^{\rm strain} = +0.4\%$.  The right panel shows a magnified band structure near $k = 0$.} \label{band}
\end{figure}

We assumed that our samples only have the epitaxial biaxial tensile strain (see Section~\ref{sec:Exp}), which breaks the crystallographic cubic symmetry due to the lattice mismatch between the GaMnAs epilayer and the relaxed GaInAs buffer layer.  The same strain parameters as for GaAs\cite{VurgaftmanetAl01JAP} were used for the GaMnAs epilayer except for its lattice constant.  Depending on the Mn ($2\sim5\%$) and In (9.5\%) contents, the strain tensor
$\varepsilon_{ii}^{\rm strain}$ $(i=x,y,z)$ varies from 0.4 to 0.57\% for the MnAs lattice constant 5.98~\AA.\cite{Ohno96.363}  We assumed that our experimental post-growth procedure does not produce any additional significant strain.  An example of the band structure calculated with the 30-band $\bf k\cdot p$ method for 5\% Mn is shown in
Fig.~\ref{band}.
%

\subsection{Hole Densities and Magnetic Anisotropy}

Before proceeding with the calculation of the frequency-dependent dielectric tensor and magneto-optical Kerr angle spectra, we roughly estimate the hole densities and Curie temperatures in our samples by calculating temperature-dependent magnetization based on the mean-field Zener model\cite{Dietletal00Science, Dietl2001}.  In addition, we determine the easy axis of magnetization with the estimated hole densities obtained for various Mn fractions. These will provide not only guidelines for the subsequent Kerr angle calculation but also justification for the results obtained by the extended $\bf k\cdot p$ method in the metallic regime in the interband range.  To
calculate the ferromagnetic critical temperature, temperature-dependent magnetization, and magnetic anisotropy, we use the 6-band $\bf k\cdot p$ method since these quantities depend only on the very top portion of the valence band structure.

\begin{figure}
\includegraphics [scale=0.57] {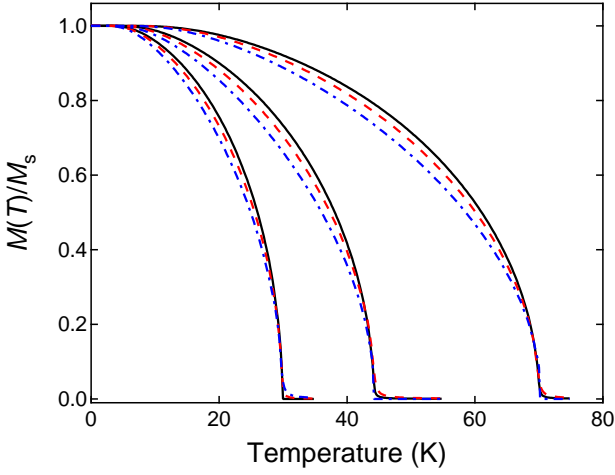}
\caption{ (Color online) Calculated temperature-dependent normalized magnetization based on the mean-field Zener model and the 6-band $\bf k\cdot p$ method with different Mn fractions $x = 0.03$ (black solid line), 0.04 (red dashed line), 0.05 (blue dash-dotted line) for $T_{\rm c} = 45$~K (sample S-2) and 70~K (annealed sample, S-3), and with $x = 0.02$ (black solid line), 0.03 (red dashed line), 0.04 (blue dash-dotted line) for $T_{\rm c} = 30$~K (sample S-1). The corresponding hole densities for each case are summarized in Table~\ref{densities}.  The results are all for the tensile strain $\varepsilon_{xx}^{\rm strain} = (0.57, 0.52, 0.46, 0.4)\%$ for Mn fraction $x = (0.02, 0.03, 0.04, 0.05)$.} \label{1theor}
\end{figure}

%
\begin{table}
\begin{center}
\begin{tabular}{c|c c}
\hline\hline
$T_{\rm c} (K)$& Mn (\%)& hole density, $p$ ($10^{20}$~cm$^{-3}$)\\
\hline
         & 3 & 2.35 \\
45 (S-2) & 4 & 1.6 \\
         & 5 & 1.2 \\
\hline
         & 3 & 5 \\
70 (S-3) & 4 & 3.05 \\
         & 5 & 2.2 \\
\hline
         & 2 & 2.4 \\
30 (S-1) & 3 & 1.39 \\
         & 4 & 0.97 \\
\hline\hline
\end{tabular}
\end{center}
\caption{Hole densities calculated for various Mn fractions based on the 6-band $\bf k\cdot p$ method and the mean-field Zener model for the measured Curie temperatures of our samples.}
\label{densities}
\end{table}
%
The temperature-dependent magnetization in Fig.~\ref{1theor} was calculated in the limit of strong degeneracy with zero external magnetic field based on the mean-field Zener model.\cite{Dietl2001}  Table~\ref{densities} lists the hole densities for various nominal Mn fractions, for which the calculated Curie temperatures are equal to the experimental ferromagnetic phase transition temperatures obtained from the SQUID measurement of the temperature-dependent magnetization in sample S-2 (Fig.~\ref{SQUID}) and the measurement of the saturated remanent Kerr angle as a function of temperature (Fig.~\ref{Anneal}).  The calculation neglects the Fermi-liquid effect, which increases $T_{\rm c}$,\cite{Dietl2001} as well as the spin-wave excitations,\cite{Konig} which decrease $T_{\rm c}$ for a given Mn fraction and a hole density.  We simulate the annealing effect,\cite{EdmondsPRL2004} which increased $T_{\rm c}$ to 70~K for sample S-3, by increasing the hole density for each Mn fraction.  Note that generally not only the hole density increases, but also the lattice constant of GaMnAs is reduced upon
annealing.\cite{Zhao}
%

\begin{figure}
\includegraphics [scale=0.57] {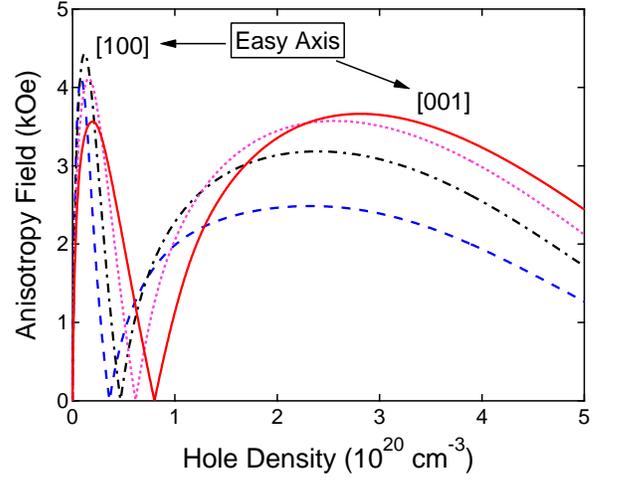}
\caption{(Color online) Anisotropy field, $H_{\rm an}$, calculated with the 6-band $\bf k\cdot p$ model, as a function of hole density under tensile strain $\varepsilon_{xx}^{\rm strain}=(0.57, 0.52, 0.46, 0.4)\%$ for Mn fractions $x=(0.02, 0.03, 0.04, 0.05)$, respectively. The magnetic easy axes of our samples are out-of-plane ([001]) for the hole densities shown in Table~\ref{densities}. The curves are for Mn fractions $x = 0.02$ (dashed), 0.03 (dash-dotted), 0.04 (dotted), and 0.05 (solid).} \label{2theor}
\end{figure}

The magneto-crystalline anisotropy, caused by the spin-orbit interaction, is attributed to the valence band holes since the total angular momentum of the local Mn moments is solely due to spins.  The anisotropy gives certain preferred directions for the spins to be aligned in crystals, so that the easy direction of magnetization is determined.  We estimated the anisotropy field, which is proportional to the difference of hole free energies in [001] and [100] directions,\cite{Dietl2001} for different nominal Mn fractions, showing magnetic easy axes under tensile strain as a function of hole density as shown in Fig.~\ref{2theor}. All hole densities shown in Table~\ref{densities} fall into the out-of-plane magnetic easy axis case ([001]) because they are all larger than the critical hole density for each Mn fraction, in which the easy axis is changed from [100] to [001].  The out-of-plane easy axis is further confirmed by the single domain like-behavior in the hysteresis loops of the measured remanent Kerr angles for the applied out-of-plane external magnetic field shown in Fig.~\ref{Anneal}.
%

\subsection{Frequency-Dependent Dielectric Tensor and Kerr Angle Spectra} \label{dielectric}

With the estimated hole densities and magnetic easy axis directions, calculations of the
dielectric tensor were performed based on the linear response theory (or the Kubo formula\cite{Marder, Yang, Sinove2003}) shown in Eq.~(\ref{eq2}) for several different hole densities (shown in Table~\ref{table3} from the range of the hole densities and Mn fractions calculated for our samples in the previous section (see  Table~\ref{densities})):
\begin{multline}\label{eq2}
\varepsilon_{\alpha\beta}(\omega) = \delta_{\alpha \beta} +
\frac{e^{2}\hbar^{2}}{m_{0}^{2}\varepsilon_{0}}\frac{1}{(2\pi)^{3}}
\sum_{\substack{a,b \\ a\neq b}}\iiint\limits_{BZ} f_{a} \frac{1}{(E_{ba})^2} \times \\
\left[\frac{p_{ab}^{\alpha} p_{ba}^{\beta}}{(E_{ba} - i\hbar\gamma -
\hbar\omega)} + \frac{p_{ba}^{\alpha} p_{ab}^{\beta}}{(E_{ba} +
i\hbar\gamma + \hbar\omega)}\right]
\mathrm{d}k_x\mathrm{d}k_y\mathrm{d}k_z
\end{multline}
where $\alpha$ and $\beta$ denote coordinates and $a$ and $b$ represent valence bands and conduction bands, respectively. Only interband transitions are included in calculations; $f_a$ is the Fermi-Dirac distribution and $p_{ab}^{\alpha}$ are components of the momentum matrix element . The integration is extended over the first Brillouin zone using full band structure calculated with the 30-band ${\bf k\cdot p}$ method described in Section \ref{band-structure}. An example of the calculated dielectric
tensor is shown in Fig.~\ref{dielectric}.
%
\begin{figure}
\includegraphics [scale=0.565] {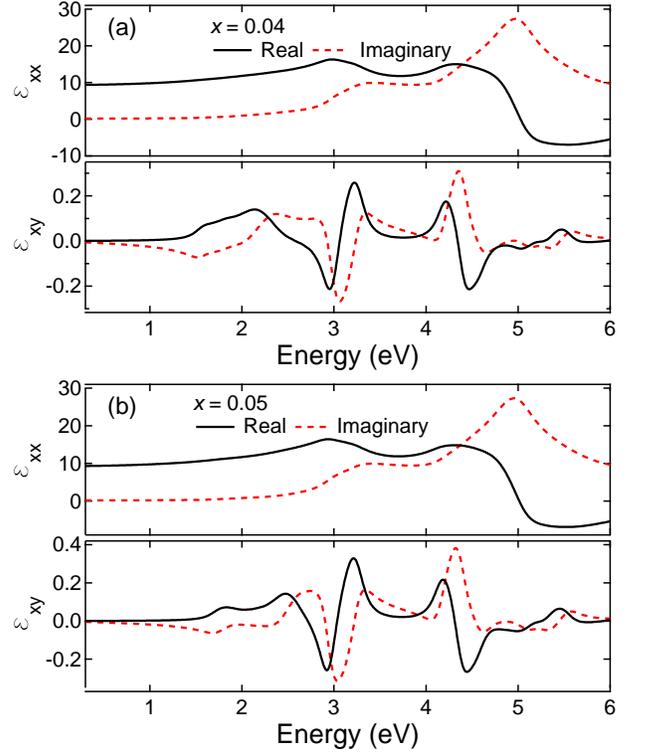}
\caption{(Color online) Complex dielectric functions $\varepsilon_{xx}$ and $\varepsilon_{xy}$ for 4\% Mn,
$p = 1 \times 10^{20}$~cm$^{-3}$ and 5\% Mn, $p = 1.5 \times 10^{20}$~cm$^{-3}$ under tensile biaxial strain.}
\label{dielectric}
\end{figure}

\begin{figure}
\includegraphics [scale=0.55] {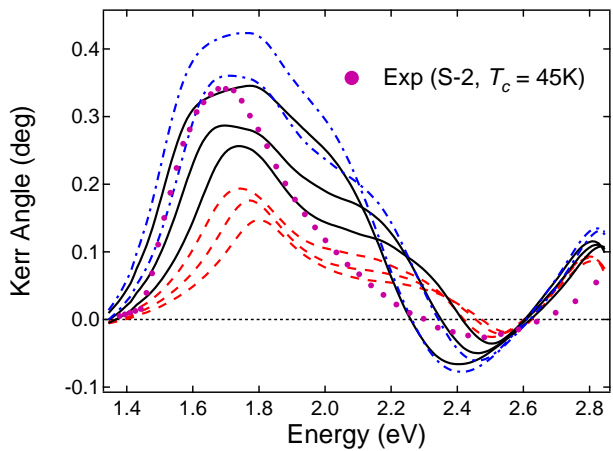}
\caption{(Color online) Calculated frequency-dependent magneto-optical Kerr angle under tensile strain by the 30-band ${\bf k\cdot p}$ method in the interband transition range for various hole densities and three different Mn fractions
$x_{\rm Mn} = 0.03$ (red dashed), 0.04 (black solid), 0.05 (blue dash-dotted) with GaMnAs epilayer thickness 70~nm. Hole densities (see Table~\ref{table3}) were chosen with the range of $\sim\pm5 \times 10^{19}$~cm$^{-3}$ around the hole density corresponding to $T_{\rm c} = 45$~K (sample S-2). For each Mn fraction, the higher peak corresponds to a lower hole density.}
\label{3theor}
\end{figure}

Once frequency-dependent dielectric functions are obtained, calculating Kerr angle spectra is straightforward, taking into account  all effects of propagation and reflections in a thin GaMnAs layer and the buffer layer.  For the dielectric function of the GaInAs buffer layer, the linear interpolation of the experimental data\cite{OptHandbook} for GaAs and InAs was used.  The calculated Kerr angle results are shown in Fig.~\ref{3theor} for sample S-2. Note that for a fixed Mn fraction, the Kerr angle peak magnitude decreases with increasing hole density. This happens because at these densities the Fermi level is already located below the lowest band edge of the spin-split $\Gamma_{\rm 8V}$.  With increasing hole density, the Fermi level is shifted further downward, which reduces the number of states that contribute to interband optical transitions.  On the other hand, if the Fermi level were initially in the middle of the spin-split $\Gamma_{\rm 8V}$, increasing the hole density would not necessarily decrease the Kerr angle amplitude.
%
%
\begin{table}
\begin{center}
\begin{tabular}{c|c|c}
\hline\hline
$T_{\rm c} (K)$& Mn (\%)&\hspace{1em}hole density, $p$ ($10^{20}$~cm$^{-3}$)\\
\hline
\multirow{8}{*}{45}   &   & 2   \\
   & 3 & 2.5 \\
   &   & 3   \\
   \cline{2-3}
   &   & 1   \\
   & 4 & 1.5 \\
   &   & 2   \\
   \cline{2-3}
   & \multirow{2}{*}{5} & 1   \\
   &   & 1.5 \\
\hline
   &   & 2   \\
   & 2 & 2.5 \\
   &   & 3   \\
   \cline{2-3}
   &   & 1.2 \\
30 & 3 & 1.4 \\
   &   & 1.6 \\
   \cline{2-3}
   &   & 0.8 \\
   & 4 & 1   \\
   &   & 1.2 \\
\hline
   &   & 2.5 \\
   & 4 & 3   \\
70 &   & 3.5 \\
   \cline{2-3}
   & \multirow{2}{*}{5} & 2   \\
   &   & 2.5 \\
\hline\hline
\end{tabular}
\end{center}
\caption{Chosen Mn fractions and corresponding hole densities for $T_{\rm c} = 45$~K (sample S-2), $T_{\rm c} = 30$~K (S-1), and $T_{\rm c} = 70$~K (S-3) used in the dielectric function and Kerr angle calculations by the 30 band ${\bf k\cdot p}$ method. The resulting Kerr angle spectra are shown in Fig.~\ref{3theor}, \ref{5theor}, and \ref{7theor}, respectively.} \label{table3}
\end{table}

Kerr angle calculation results for 4\% Mn give the best agreement with the experimental MOKE spectra of S-2, as shown in Fig.~\ref{3theor}. The first positive peak that can be attributed to the interband transitions around the $E_0$ critical point is red-shifted due to phenomenological band gap narrowing of 0.121 and 0.138~eV, respectively, for each case.  Note that the above Mn fractions are 1.5-2 times larger than the experimental nominal value, 2.4\% for S-2, as we discuss in the following section.

\begin{figure}
\includegraphics [scale=0.55] {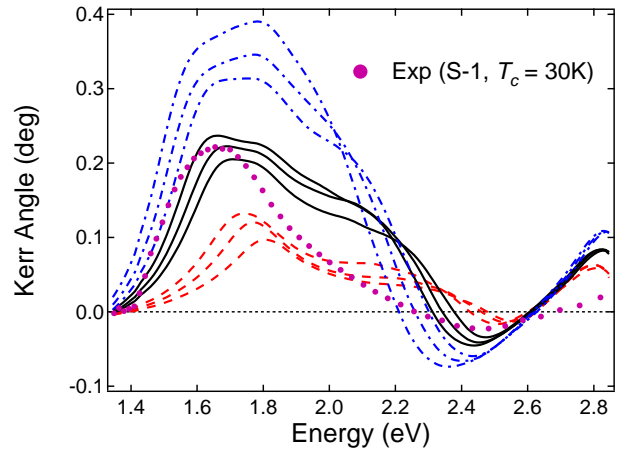}
\caption{ (Color online) Frequency-dependent magneto-optical Kerr angle under tensile strain calculated through by the 30-band ${\bf k\cdot p}$ method in the interband transition range for various hole densities and three different Mn fractions $x_{\rm Mn} = 0.02$ (red dashed), 0.03 (black solid), 0.04 (blue dash-dotted) with GaMnAs epilayer thickness of 70~nm. Hole densities (Table~\ref{table3}) were chosen around the estimated hole densities at $T_{\rm c} = 30$~K from Table~\ref{densities} (sample S-1). For each Mn fraction, corresponding hole densities are smaller from the top.}
\label{5theor}
\end{figure}

For sample S-1, the calculated result for Kerr angles is closest to the experimental MOKE spectra for 3\% Mn, as shown in Fig.~\ref{5theor} and Table~\ref{table3}. This Mn fraction is also larger than the experimental value of 1.5\% for sample S-1.
%

\begin{figure}
\includegraphics [scale=0.55] {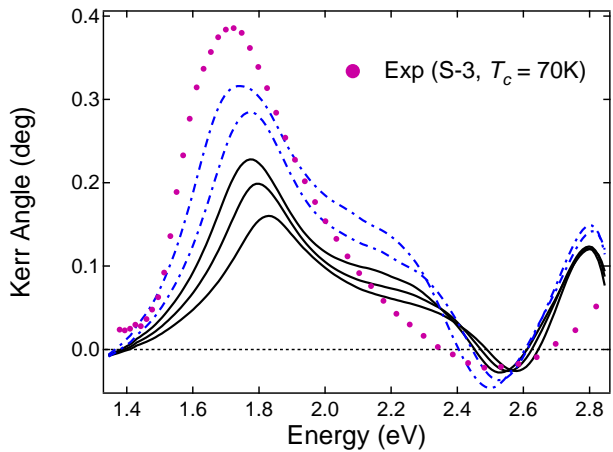}
\caption{ (Color online) Frequency-dependent magneto-optical Kerr angle calculated through the 30-band ${\bf k\cdot p}$ method in the interband transition range for various hole densities and two different Mn fractions $x_{\rm Mn} = 0.04$ (black solid), 0.05 (blue dash-dotted) with GaMnAs epilayer thickness of 70~nm. Hole densities (shown in Table~\ref{table3}) were chosen around the estimated hole densities (Table~\ref{densities}) for $T_{\rm c} = 70$~K (sample S-3) obtained by the 6-band ${\bf k\cdot p}$ method and the mean-field Zener model. For each Mn fraction, corresponding hole densities are smaller from the top.} \label{7theor}
\end{figure}

In Table~\ref{densities}, the hole densities that are predicted for $T_{\rm c} = 70$~K based on the 6-band ${\bf k\cdot p}$ model are overestimated since at large hole densities the hole free energy calculated with the 6-band ${\bf k\cdot p}$ model becomes non-negligibly smaller than the one calculated with the full-band model. This originates from the difference between the valence band structures calculated with these two models at large $k$. Therefore, it is expected that the full band structure calculation yields the hole densities lower than those listed in Table~\ref{densities}.  Note also that with 4\% or 5\% Mn fraction, the Kerr angle calculation agrees better with the experimental Kerr spectra for $T_{\rm c} = 45$~K (S-2), as shown in Fig.~\ref{3theor}. Therefore, for the annealed case of $T_{\rm c} = 70$~K the Kerr angle calculations are performed only for 4\% and 5\% Mn fractions. The results are shown in Fig.~\ref{7theor}, and their corresponding hole densities are listed in Table~\ref{table3}.
%

\section{Discussion}

Our calculated MOKE spectra of ferromagnetic GaMnAs samples in the interband transition range reveal a general pattern of a large-amplitude positive peak around 1.4-2.3~eV followed by negative and positive peaks with lower-amplitudes at higher photon energies, similar to the experimental data.  At the same time, the quantitative spectral shapes and the positions of the peaks sensitively depend on the amount of substitutional Mn, the hole density, and the layer thickness.  We determined the range of Mn fractions and corresponding hole densities for which the calculated Curie temperatures were equal to the experimentally measured ferromagnetic transition temperatures. Then the parameters from this range providing the best fit to the measured MOKE spectra were found.  The resulting Mn
fractions turned out to be larger than the experimental nominal doping values for all samples.  If we assume that the experimental Mn concentrations are quite accurate, this result could indicate that antiferromagnetic $p$-$d$ exchange coupling strengths are 1.5-2 times stronger than the value of $J_{ij}$ adopted in the simulations.  This is because in the mean-field approximation that we used throughout the calculations the exchange coupling Hamiltonian contains the product of $N_{\rm Mn}$ and $J_{ij}$, as shown in Eq.~(\ref{eq1}).
%
\begin{figure}
\includegraphics [scale=0.55] {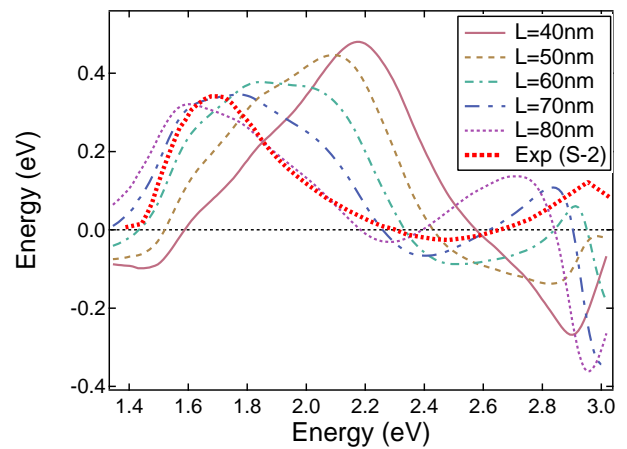}
\caption{(Color online) Calculated MOKE spectra for ferromagnetic GaMnAs with different layer thicknesses for 4\% Mn fraction and hole density of $1 \times 10^{20}$~cm$^{-3}$.  The dotted line shows the experimental spectrum for sample S-2.} \label{thickness}
\end{figure}

Unlike magnetic circular dichroism, which has a clear physical interpretation as the difference of absorption coefficients between the $\sigma^-$ and $\sigma^+$ polarizations, it is difficult to extract a single physical parameter that determines the characteristics of the Kerr angle spectra in a thin layer limit\cite{LangetAl2005PRB} since the contribution of the GaMnAs layer thickness, diagonal and off-diagonal components of dielectric functions, and the dielectric function of the buffer layer should be explicitly considered.  However, the first positive peak can still be attributed to the interband transitions around the $E_0$ critical point.  The position of this peak and those of subsequent peaks are affected by the layer thickness.  Figure~\ref{thickness} shows a set of calculated MOKE spectra for a fixed Mn fraction (4\%) and hole density ($p = 1 \times 10^{20}$~cm$^{-3}$) but different thicknesses of the GaMnAs epilayer.  The Kerr angle peaks red-shift as the thickness increases.  The best match to the spectrum of sample S-2 is obtained when the layer thickness is about 70~nm, which is somewhat higher than our nominal value of 50~nm.

For the annealed sample (S-3) the main peak in the measured MOKE spectrum shows a slightly increased amplitude as well as a slight blue-shift as compared to that before annealing (S-2).  If we assume that the only effect of annealing is an increased hole density for a given Mn fraction, our simulations predict an opposite trend: a decrease in the amplitude of the Kerr angle peak with increasing hole density.  Within the mean field approximation, the observed effect of annealing indicates that the annealing also leads to an increase in the exchange interaction energy through an  increase in the Mn fraction or/and $J_{ij}$.

\section{Conclusions}

We have used the magneto-optical Kerr effect to investigate three GaMnAs samples with out-of-plane anisotropy but different Mn densities.  We measured the magneto-optical Kerr rotation as a function of three continuously varying parameters: photon energy, magnetic field, and temperature.  We observed remanent Kerr spectral peaks near 1.7~eV, which increased in intensity and blue-shifted with Mn-doping and further blue-shifted with annealing.  We presented
theoretical modeling of magnetic and optical properties of ferromagnetic GaMnAs using the mean-field approximation and the 30-band ${\bf k\cdot p}$ method, in which interband optical transitions are calculated over the whole first Brillouin zone and modified valence band interactions with remote bands affected by the Mn-hole spin exchange interaction, the biaxial strain, and the phenomenological Coulomb interaction and disorder effect are explicitly and simultaneously taken into account. Our calculation of MOKE spectra of thin-film ferromagnetic GaMnAs samples in the metallic regime is in good agreement with the experimental spectra.  This supports the GaMnAs electronic
band structure in which the impurity band is smeared out into the valence band, although it cannot rule out the presence of impurity states in the band gap.  To treat the MOKE spectra more rigorously in the interband range within the full band ${\bf k\cdot p}$ method, the $k$-dependent self-energy, which explicitly takes into account many-body Coulomb interaction and disorder effects, has to be calculated.

\begin{acknowledgments}
Y.-H. Cho and A. Belyanin acknowledge many helpful discussions with J. Sinova and T. Jungwirth.  This work was supported by the NSF through Award Nos.~OISE-0530220 and ECS-0547019.
\end{acknowledgments}


\end{document}